\documentclass[aps, prd,reprint,groupedaddress,amsmath,amssymb,nolongbibliography, floatfix, nofootinbib]{revtex4-2}

\usepackage{bm}
\usepackage{booktabs}
\usepackage{graphicx}
\usepackage{xcolor}
\usepackage{hyperref}

\hypersetup{
 hidelinks,
 pdftitle={Two-loop correction to the complex-bion amplitude in a quasi-exactly solvable sextic oscillator},
 pdfauthor={Adrian M. Escobar-Ruiz}}
\graphicspath{{./}}

\newcommand{\dd}{\mathrm d}
\newcommand{\cO}{\mathcal O}
\newcommand{\cH}{\mathcal H}

\newcommand{\cC}{\mathcal C}

\newcommand{\med}{\operatorname{med}}

\begin{document}

\title{Two-loop correction to the complex-bion amplitude in a quasi-exactly solvable sextic oscillator}

\author{Adri\'an M. Escobar-Ruiz}
\email[Contact author: ]{admau@xanum.uam.mx}
\affiliation{Departamento de F\'isica, Universidad Aut\'onoma Metropolitana--Iztapalapa, San Rafael Atlixco 186, 09340 Ciudad de M\'exico, Mexico}

\begin{abstract}
In this study, a quasi-exactly solvable sextic potential is analyzed by two diagrammatic calculations. At positive coupling, factorization fixes the ground-state energy $E_0$ exactly and forces every perturbative coefficient to vanish. Using only Feynman diagrams, we verify this cancellation explicitly through third order. After continuation to negative coupling, the formal zero mode becomes nonnormalizable. A real instanton remains, but its one-instanton contribution to \(E_0\) vanishes. In particular, the leading shift in energy arises from a complex neutral bion. Taking an exact constrained family, we calculate its local median amplitude through two loops. The direct diagrammatic sum gives $c_{\rm 2L}^{\rm diag}=-0.750000(5)$, in excellent agreement with the exact-WKB value $-3/4$.
\end{abstract}

\maketitle

\section{Introduction}

Finite-action Euclidean saddles control tunneling amplitudes and exponentially small spectral effects \cite{Coleman}. Correlated instanton sectors also govern the large-order behavior and ambiguity
structure of perturbation theory, providing the nonperturbative sectors of the transseries
\cite{BogomolnyIbarI,ZinnJustinMulti,VerbaarschotWu,DunneUnsalZero}.

Exact spectral protection and nontrivial semiclassical saddles coexist in the sextic potential studied here. At positive coupling, factorization produces a normalizable ground state and fixes its energy. Interestingly, every coupling-dependent coefficient in its perturbative expansion vanishes (see also \cite{ShuryakComplexVacuum2026}). After analytical continuation to negative coupling, the factorization identity survives, whereas the continued zero mode grows at infinity and leaves the physical Hilbert space. In this case, the ground-state energy may receive an exponentially small correction \cite{DunneUnsalZero}.

The continued sextic potential has a parity-symmetric triple-well profile in the weak-coupling limit. A real instanton connects the central minimum to either of the two equivalent outer minima. Because the endpoint curvatures are unequal, the relevant one-instanton amplitude vanishes. Consequently, the leading
nonperturbative contribution to the ground-state energy comes from a neutral bion, an instanton--anti-instanton configuration that leaves the central well and returns to it \cite{DunneTriple,BehtashCritical}. The instanton and anti-instanton pair can separate almost freely. In particular, integrating over this separation along the two lateral contours produces conjugate
complex-bion saddles, and their median defines the local one-bion amplitude
\cite{BehtashComplex,BehtashCritical,BehtashKamata2026}.

Explicitly, we consider the 1d QM Hamiltonian 
\begin{equation}
 \begin{aligned}
& H=-\frac12\frac{\dd^2}{\dd x^2}
 +\frac12\left[\nu^2x^6+2\mu\nu x^4+(\mu^2-3\nu)x^2\right],
 \\ & \nu\in\mathbb{R},\quad \mu>0 ,
 \end{aligned}
 \label{eq:H}
\end{equation}
with atomic units $\hbar=1$, $m=1$. This Hamiltonian operator, our main object of study, is the lowest algebraic member of the standard quasi-exactly solvable sextic family \cite{TurbinerUshveridze,TurbinerCMP}.

Using only the tools from quantum field theory, we perform two complementary diagrammatic calculations. In the perturbative sector with $\nu>0$, we verify the cancellation of perturbative contributions by summing the Feynman diagrams through $O(\nu^3)$. After setting $\nu=-g$, an exact constrained
family provides the Green function, reduced determinant, collective-coordinate measure, and the regulated two-loop bion sum. The resulting two-loop (2L) coefficient,
\[
c_{\rm 2L}^{\rm diag}=-0.750000(5),
\]
agrees with the exact-WKB value $-3/4$. 

The underlying triple-well mechanism was established by Dunne, Sulejmanpasic, and \"Unsal \cite{DunneTriple}, and the associated exact-WKB quantization conditions were derived by Kamata
\emph{et al.} \cite{KamataExactWKB}. Behtash and Kamata recently related the lateral quasi-zero-mode cycles to exact-WKB Stokes jumps at dilute one-loop order \cite{BehtashKamata2026}. Here, the local
path-integral calculation is extended to two loops for the sextic complex bion.

The present calculation is motivated by the diagrammatic program initiated by Aleinikov and Shuryak for loop corrections around instanton backgrounds \cite{AleinikovShuryak}. It was developed further by Olejn\'ik and by W\"ohler and Shuryak \cite{Olejnik,ShuryakWohler}, and extended to three loops by Escobar-Ruiz, Shuryak, and Turbiner \cite{EscobarQuartic,EscobarSine}. Its relation to Shuryak's recent
quartic potential is direct \cite{ShuryakComplexVacuum2026}: both models have a vanishing shifted perturbative series and acquire nonperturbative lifting when the formal zero mode becomes
nonnormalizable \cite{DunneUnsalZero,BehtashComplex,ShuryakComplexVacuum2026}. Shuryak's asymmetric quartic potential is not a parameter specialization of the present sextic model, although both belong to
the same factorized class and exhibit the same perturbative-protection mechanism. In \cite{ShuryakComplexVacuum2026}, the spectrum and complex trajectories were found numerically. Here, the complex background and collective-coordinate reduction are obtained analytically, while the two-loop graph contractions are evaluated numerically with controlled continuum and infinite-time extrapolations.

Of course, the constrained-family and collective-coordinate formalism is standard \cite{AffleckConstrained,AoyamaValleysQM,BehtashCritical}. Our specific advance is a direct diagrammatic construction of the local one-bion amplitude for the quasi-exactly solvable sextic potential, together with an explicit perturbative cancellation through third order. Exact WKB independently checks the two-loop coefficient and fixes the global spectral multiplier in the present normalization \cite{KamataExactWKB}.

The structure of the paper is simple. Section~\ref{sec:pert} treats the perturbative cancellation. Sec.~\ref{sec:bion} develops the complex-bion expansion and exact-WKB comparison, and finally Sec.~\ref{conclusions} summarizes the results. Technical derivations are collected in the
appendices.

\section{Perturbative diagrammatic cancellation}
\label{sec:pert}

For the Hamiltonian (\ref{eq:H}), factorization fixes the ground-state energy before any perturbative calculation. In this section, we examine how this protection emerges from cancellations among Feynman diagrams around the harmonic vacuum.

\begin{figure*}[!t]
 \centering
 \includegraphics[width=0.95\textwidth]{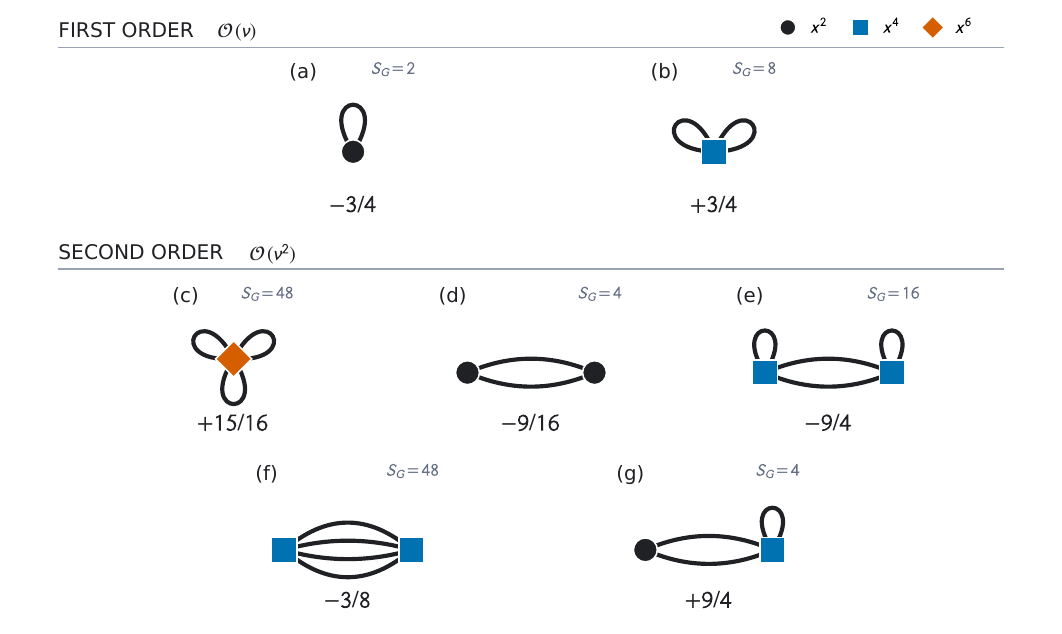}
 \caption{Connected vacuum graphs through $O(\nu^2)$. Filled circles, squares, and diamonds denote the vertices $-3\nu x^2/2$, $\nu x^4$, and $\nu^2x^6/2$, respectively. The black number below each graph gives its complete contribution to the ground-state energy. Gray labels denote the symmetry-factor denominators $S_G$, defined such that the graph carries the combinatorial weight $1/S_G$ when its vertices are factorially normalized.}
 \label{fig:P12}
\end{figure*}

\subsection{Factorization and the protected ground state}

It is convenient to introduce the prepotential $\Phi(x)$ and the associated first-order operators
\begin{equation}
 \begin{aligned}
 \Phi(x)&=\frac{\mu x^2}{2}+\frac{\nu x^4}{4},\\
 A&=\frac{\dd}{\dd x}+\Phi'(x),
 &A^\dagger&=-\frac{\dd}{\dd x}+\Phi'(x).
 \end{aligned}
 \label{eq:factoroperators}
\end{equation}
Since $\Phi'(x)=\mu x+\nu x^3$, the Hamiltonian~\eqref{eq:H} takes the factorized form \cite{VerbaarschotWu}
\begin{equation}
 H=\frac12 A^\dagger A+\frac{\mu}{2}.
 \label{eq:factor}
\end{equation}
For $\mu>0$ and $\nu>0$, the equation $A\psi_0=0$ gives the ground state
\begin{equation}
 \begin{gathered}
 \psi_0(x)=\mathcal N(\mu,\nu)
 \exp\!\left(-\frac{\mu x^2}{2}-\frac{\nu x^4}{4}\right),
 \qquad
 E_0=\frac{\mu}{2},\\[1mm]
 \mathcal N(\mu,\nu)=
 \left(\frac{2\nu}{\mu}\right)^{1/4}
 \frac{\exp[-\mu^2/(8\nu)]}
 {\sqrt{K_{1/4}\!\left(\mu^2/(4\nu)\right)}} \,,
 \end{gathered}
 \label{eq:exactstate}
\end{equation}
where $K_{1/4}$ is the modified Bessel function of the second kind. 

Because the eigenvalue in \eqref{eq:exactstate} is independent of $\nu$, every coefficient in the perturbative expansion of $E_0-\mu/2$ vanishes identically. This formal series remains zero after continuation to $\nu<0$, although the continued solution of $A\psi=0$ no longer belongs to $L^2(\mathbb R)$. At positive coupling, the exact eigenvalue also constrains the full transseries: the combined nonperturbative contribution to this level must vanish when the analytically continued saddle amplitudes are assembled \cite{DunneUnsalZero}.

It is worth mentioning that a separate geometric regime occurs for $\nu/\mu^2>1/3$. The positive-coupling potential then develops two parity-related minima and supports a standard real instanton, but this double-well geometry disappears before the weak-coupling limit $\nu/\mu^2\to0$ is reached. Throughout the positive-coupling domain, the form of ~\eqref{eq:factor} keeps the algebraic ground-state energy fixed at $\mu/2$.

So, factorization proves the perturbative protection. We now verify this cancellation directly from Feynman diagrams. With $t$ denoting real time, Wick rotate and introduce the
dimensionless variables
\[
 q=\sqrt{\mu}\,x,\qquad
 \lambda=\frac{\nu}{\mu^2},\qquad
 \widehat H=\frac{H}{\mu},\qquad
 \tau=i\mu t\,,
\]
$\tau$ being the dimensionless Euclidean time. For the remainder of this section, we
relabel $(q,\lambda,\widehat H,E_0/\mu)\to(x,\nu,H,E_0)$. After this rescaling, the Hamiltonian~(\ref{eq:factor}) becomes
\begin{equation}
 H=H_{\rm HO}+\nu V_1+\nu^2V_2,
 \qquad
 H_{\rm HO}=-\frac12\frac{\dd^2}{\dd x^2}+\frac{x^2}{2},
 \label{eq:verticesP}
\end{equation}
with ground state energy $E_0=\frac{1}{2}$ and
\begin{equation}
 V_1(x)=x^4-\frac32x^2,
 \qquad
 V_2(x)=\frac12x^6.
 \label{eq:interactionvertices}
\end{equation}
The monomials in $V_1$ and $V_2$ generate the quadratic, quartic, and
sextic vertices of the model.

The familiar harmonic Green function is
\begin{equation}
 G_0(\tau-\tau')
 =\frac12e^{-|\tau-\tau'|}.
 \label{eq:harmonicpropagator}
\end{equation}
Denoting by $\mathcal T$ the dimensionless Euclidean-time extent, we write
\begin{equation}
 \mathcal I_j[x]=
 \int_{-\mathcal T/2}^{\mathcal T/2}
 \dd\tau\,V_j\!\left(x(\tau)\right),
 \qquad j=1,2,
 \label{eq:integratedvertices}
\end{equation}
so that the interaction part of the Euclidean action takes the form
\[
 S_{\rm int}[x]=\nu\,\mathcal I_1[x]+\nu^2\,\mathcal I_2[x].
\]
The normalized vacuum amplitude behaves at large $\mathcal T$ as
\[
 \left\langle e^{-S_{\rm int}}\right\rangle_0
 \sim
 e^{-\mathcal T(E_0-1/2)}\,,
\]
and
\begin{equation}
 E_0-\frac12
 =-\lim_{\mathcal T\to\infty}\frac1{\mathcal T}
 \ln\left\langle e^{-S_{\rm int}}\right\rangle_0 .
 \label{eq:vacuumfunctional}
\end{equation}

Expanding through $O(\nu^3)$ gives
\begin{align}
 E_0-\frac12
 &=\lim_{\mathcal T\to\infty}\frac1{\mathcal T}\Bigg\{
 \nu\langle\mathcal I_1\rangle_0
 +\nu^2\left[
 \langle\mathcal I_2\rangle_0
 -\frac12\langle\mathcal I_1^2\rangle_{0,c}
 \right]
 \nonumber\\
 &+\nu^3\left[
 -\langle\mathcal I_1\mathcal I_2\rangle_{0,c}
 +\frac16\langle\mathcal I_1^3\rangle_{0,c}
 \right]\Bigg\}
 +O(\nu^4).
 \label{eq:cumulant}
\end{align}
Here $\langle\cdots\rangle_0$ denotes the harmonic-vacuum expectation
value. Euclidean time ordering is implicit, and the subscript $c$
selects the connected part.

\subsection{Results. First three orders}

For the Hamiltonian (\ref{eq:verticesP}), the shifted ground-state energy as a power series in $\nu$ is
\[
 E_0-\frac12=\sum_{n\geq1}\nu^n E_0^{(n)}.
\]
Figure~\ref{fig:P12} displays the Feynman diagrams involved in the cancellations at the first two orders. For instance, at first order only two diagrams occur and 
\begin{equation}
 E_0^{(1)}=-\frac34+\frac34=0.
 \label{eq:P1}
\end{equation}
At second order, with five contributions shown in panels (c)--(g), we have
\begin{equation}
 E_0^{(2)}
 =\frac{15}{16}-\frac9{16}-\frac94
 -\frac38+\frac94=0.
 \label{eq:P2}
\end{equation}

\begin{figure*}[t]
 \centering
 \includegraphics[width=0.95\textwidth]{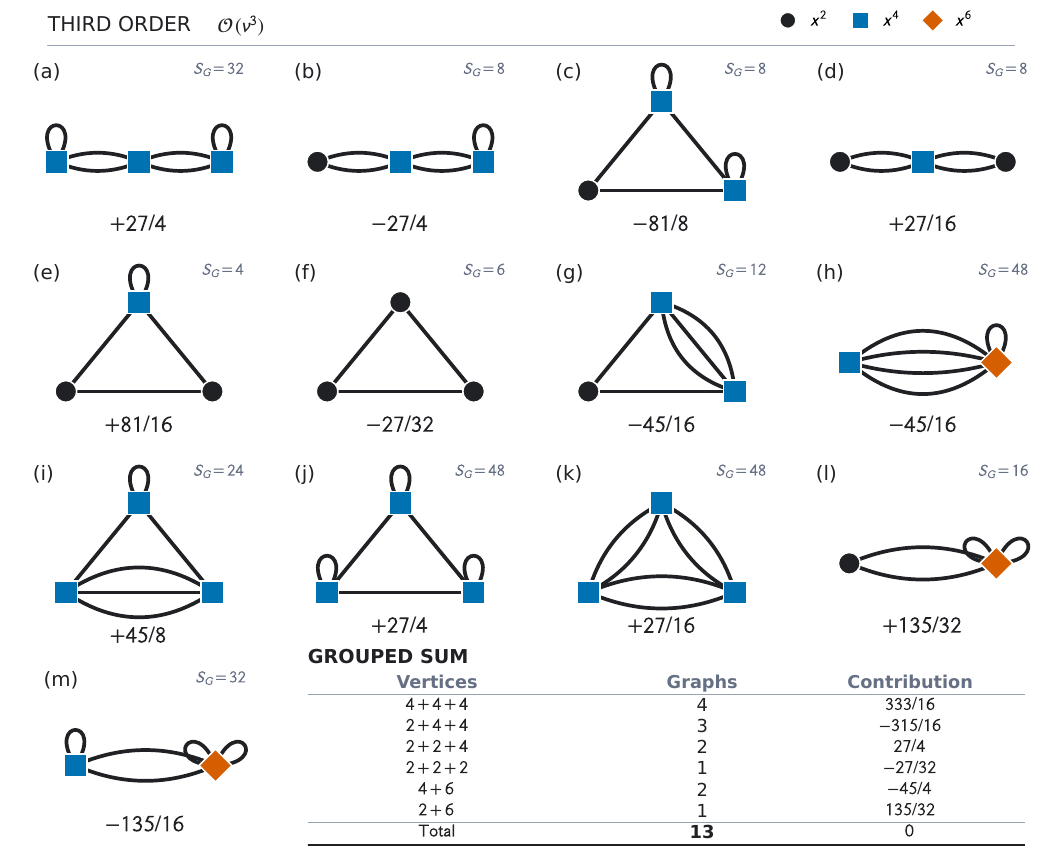}
 \caption{The thirteen connected vacuum graphs at order $\nu^3$. Black numbers give their complete contributions to $E_0^{(3)}$. Gray labels denote the symmetry-factor denominators $S_G$, defined such that the graph carries the combinatorial weight $1/S_G$ when its vertices are factorially normalized. The table collects the graphs by vertex content; the total is zero.}
 \label{fig:P3}
\end{figure*}

At order $\nu^3$, thirteen diagrams are generated from
$\mathcal I_1^3$ and $\mathcal I_1\mathcal I_2$.
Their individual values are shown in Fig.~\ref{fig:P3}. As expected, the complete sum again vanishes (we actually checked by pure diagrammatic means this cancellation up to order $\nu^4$).

For pedagogical reasons, description of one graph illustrates these simple calculations. Figure~\ref{fig:P3}(c) contains one quadratic vertex, two quartic vertices, two self-contractions, and three lines joining the vertices. The factorially normalized vertex coefficients are $-3$ and $4!$, while its symmetry-factor $S_{(c)}=8$. Fixing the quadratic vertex at $\tau=0$ gives
\begin{equation}
 E_{0,(c)}^{(3)}
 =
 \frac{(-3)(4!)^2}{8}\,
 G_0(0)^2\,\mathcal I_{\triangle},
 \label{eq:graph-c-integral}
\end{equation}
where
\begin{align}
 \mathcal I_{\triangle}
 &\equiv
 \int_{\mathbb R}\dd\tau_1
 \int_{\mathbb R}\dd\tau_2\,
 G_0(\tau_1)G_0(\tau_2)
 \times
 G_0(\tau_1-\tau_2).
 \label{eq:triangle-integral}
\end{align}
Using $G_0(\tau)=e^{-|\tau|}/2$, this integral becomes
\begin{align}
 \mathcal I_{\triangle}
 &=
 \int_{-\infty}^{\infty}
 \frac{\dd\omega}{2\pi}\,
 \frac{1}{(\omega^2+1)^3}
=\frac{3}{16}.
 \label{eq:triangle-fourier}
\end{align}
Consequently,
\begin{equation}
 E_{0,(c)}^{(3)}
 =
 \frac{(-3)(4!)^2}{8}
 \left(\frac12\right)^2
 \frac{3}{16}
 =-\frac{81}{8},
\end{equation}
as shown in Fig.~\ref{fig:P3}(c). The remaining diagrams follow from the same rules. Factorization
requires their contributions to cancel at every order. To express
this constraint diagrammatically, the connected vacuum functional
gives
\begin{equation}
 E_0^{(n)}
 =
 \lim_{T\to\infty}\frac1T
 \sum_{b=0}^{\lfloor n/2\rfloor}
 \frac{(-1)^{\,n-b+1}}
 {(n-2b)!\,b!}
 \left\langle
 \mathcal I_1^{\,n-2b}\mathcal I_2^{\,b}
 \right\rangle_{0,c}.
 \label{eq:allorder}
\end{equation}
Here $b$ counts insertions of $\mathcal I_2$, and $n-2b$ those of $\mathcal I_1$. Factorization and the exact state in~\eqref{eq:exactstate} require
\begin{equation}
 E_0^{(n)}=0,
 \qquad n\geq1.
\end{equation}
Restoring the variables of~\eqref{eq:H} and
setting $\nu=-g<0$, the perturbative expansion of
$E_0-\mu/2$ remains identically zero, whilst the physical level is
lifted by an exponentially small neutral-bion contribution of order
$e^{-\mu^2/(2g)}$. This will be addressed below.

\section{Complex-bion lifting and its two-loop expansion}
\label{sec:bion}

In this section, we compute the local one-bion amplitude and its two-loop fluctuation
expansion using the median of the two lateral separation cycles.

\subsection{Continued triple well and exact constrained family}

In (\ref{eq:H}) we set $\nu=-g$, with $g>0$, and introduce
\begin{equation}
 q=\sqrt{\frac{g}{\mu}}\,x,
 \qquad
 h=\frac{g}{\mu^2},
 \label{eq:scaling}
\end{equation}
while retaining the dimensionless Euclidean time $\tau$ defined above.
After subtracting the algebraic level $\mu/2$, define
\begin{align}
 \mathcal H_+
 &=
 \frac{g}{\mu^3}
 \left[H(-g)-\frac{\mu}{2}\right],
 \nonumber\\
 \delta E
 &=
 \frac{E_0(-g)}{\mu}-\frac12
 =\frac{\mathcal E_+(h)}{h},
 \label{eq:scaleddefinition}
\end{align}
where $\mathcal E_+(h)$ is the lowest eigenvalue of
\begin{equation}
 \mathcal H_+
 =\frac12\left[
 -h^2\partial_q^2
 +q^2(q^2-1)^2+h(3q^2-1)
 \right]\,,
 \label{eq:Hplus}
\end{equation}
and $h$ controls its semiclassical expansion.
We denote the deformed potential in~\eqref{eq:Hplus} by
\[
 U_+(q;h)=\frac12\left[q^2(q^2-1)^2+h(3q^2-1)\right],
 \quad
 V_0(q)\equiv U_+(q;0).
\]
The vacuum-subtracted potential entering the Euclidean action is
$U_+(q;h)-U_+(0;h)=V_0(q)+3hq^2/2$.

For $0<h<1/9$, the potential has a parity-symmetric triple-well
profile. As $h\to0$, its minima approach $q=0,\pm1$, with
frequencies $\omega_{\rm c}=1$ and $\omega_{\rm o}=2$. A
central-to-outer instanton has $S_I=1/4$, but its normalized
amplitude contains
$\exp[-(\omega_{\rm o}-\omega_{\rm c})T/4]$ and vanishes as
$T\to\infty$ \cite{DunneTriple}. The leading closed configuration
is therefore a neutral pair with action $2S_I=1/2$, whose lateral
continuation reaches the conjugate complex-bion saddles
\cite{BehtashComplex,BehtashCritical,BehtashPicardLefschetz}.

Again, factorization fixes the sign of the energy shift:
\begin{equation}
 H(-g)-\frac{\mu}{2}
 =\frac12A^\dagger A\geq0.
\end{equation}
The formal zero mode grows as
$\exp[(q^4/4-q^2/2)/h]$ and is not square integrable. So,
$E_0(-g)>\mu/2$ or equivalently $\delta E>0$.

The bion center and constituent separation must be treated as
collective coordinates. The separation is nearly free, so its
integration cannot be replaced by a Gaussian fluctuation.
We compute the transverse one- and two-loop
contributions at fixed separation and perform the lateral
separation integral last.

\begin{figure*}[t]
 \centering
 \includegraphics[width=0.9\textwidth]{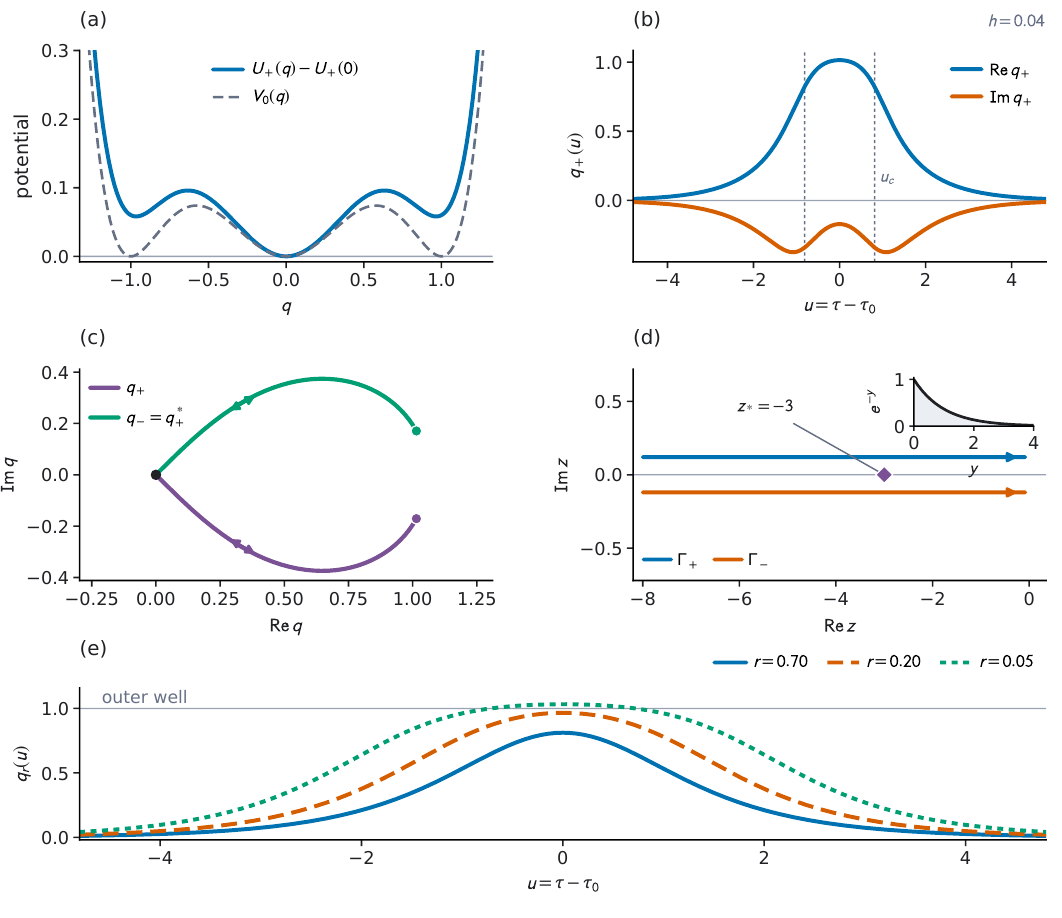}
 \caption{Complex-bion geometry at $h=0.04$. (a) Classical potential $V_0(q)$ and vacuum-subtracted deformed
potential $U_+(q;h)-U_+(0;h)$. (b) Real and imaginary parts of the saddle with $r=+i\sqrt{3h}$; vertical lines mark $|r|\cosh(2\Omega u_c)=1$. (c) Conjugate trajectories in the complex-$q$ plane. Each reaches a turning point and retraces the same arc; the curves show the classical paths, not the full configuration-space thimbles. (d) Upper and lower dilute separation cycles in $z=r^2/h$, with saddle $z_*=-3$; the inset shows the leading weight $e^{-y}$. (e) Real constrained family. As $r\to0$, the instanton and anti-instanton separate and an outer-well plateau develops.}
 \label{fig:biongeometry}
\end{figure*}

After subtracting the central-vacuum contribution, the dimensionless Euclidean action is
\begin{equation}
 s[q]=\int\dd\tau\left[
 \frac12\dot q^{\,2}
 +\frac{\Omega^2}{2}q^2-q^4+\frac12q^6
 \right],
 \qquad \Omega^2=1+3h,
 \label{eq:action}
\end{equation}
with path-integral weight $\exp[-s[q]/h]$. Following the constrained-instanton and valley framework \cite{AffleckConstrained,AoyamaValleysQM}, the separation coordinate is
introduced through the exact constrained family
\begin{equation}
 q_r^{(\sigma)}(\tau)
 =
 \sigma\frac{\Omega}
 {\sqrt{1+r\cosh[2\Omega(\tau-\tau_0)]}},
 \qquad \sigma=\pm1 .
 \label{eq:valley}
\end{equation}
Here $\tau_0$ is the center and $r$ controls the constituent
separation.  As $r\to0$, the instanton and anti-instanton separate and an outer-well plateau develops.  The branch $\sigma=+1$ describes an excursion through the right outer well, $q\simeq+1$, whereas $\sigma=-1$ gives its parity image through the left outer well, $q\simeq-1$.  For either value of $\sigma$, continuing $r$ to $r_*=\pm i\sqrt{3h}$ produces the conjugate complex-bion saddles. The above family obeys
\begin{equation}
 \left.\frac{\delta s}{\delta q}\right|_{q_r}
 =\kappa(r)q_r^5,
 \qquad
 \kappa(r)=\frac{3(\Omega^2+r^2-1)}{\Omega^2},
 \label{eq:valleyeq}
\end{equation}
so $q_r$ satisfies the constrained-stationarity condition $\delta(s-\kappa\cC)=0$, where
\begin{equation}
 \cC[q]=\frac16\int q^6\dd\tau.
\end{equation}
Let $C(r)=\mathcal C[q_r]$ and $\psi_r=\partial_rq_r$.  The constraint
normal is
\[
 \chi(\tau)\equiv
 \left.\frac{\delta\mathcal C}{\delta q(\tau)}\right|_{q_r}
 =q_r^5(\tau).
\]
With $(f,g)\equiv\int f(\tau)g(\tau)\,d\tau$, the variation along the family reads
\[
 C'(r)=(\chi,\psi_r),
 \qquad
 s'(r)=\kappa(r)C'(r)\,,
\]
and the unconstrained saddles satisfy $\kappa(r_\ast)=0$, or
\[
 r_\ast^\pm=\pm i\sqrt{3h}.
\]
For each parity branch, $q_{r_\ast^\pm}$ are conjugate complex bions
\cite{BehtashHomological,BehtashPicardLefschetz}, with actions
\begin{equation}
 s_{\rm cb}^{\pm}
 =\frac{\Omega}{2}
 +\frac{3h}{2}\ln\!\frac{\Omega+1}{\sqrt{3h}}
 \mp i\frac{3\pi h}{4}.
 \label{eq:complexaction}
\end{equation}
The weights $e^{-s_{\rm cb}^{\pm}/h}$ carry phases
$e^{\pm3\pi i/4}$.  The determinant branch and lateral orientation fix
the overall semiclassical sign
\cite{BehtashComplex,BehtashCritical}.  Figure~\ref{fig:biongeometry}
shows the continuation from the real dilute pair to these conjugate
saddles.

\subsection{Green function and one-loop determinant}

The center and separation are already represented by collective
coordinates. To avoid counting these motions twice, we integrate
only the transverse fluctuations at fixed $r$
\cite{DunneTriple,BehtashCritical}. We expand about the constrained family as $q=q_r+\sqrt h\,\eta+\cdots$. At fixed $r$, the transverse fluctuation operator is the
second variation of the constrained action
$s-\kappa(r)C$ \cite{AffleckConstrained,AoyamaValleysQM}, namely
\begin{equation}
 \cO_v=\left.\frac{\delta^2(s-\kappa\cC)}{\delta q^2}\right|_{q_r}
 =-\partial_\tau^2+\Omega^2-12q_r^2
 +15\frac{\Delta^2}{\Omega^2}q_r^4.
 \label{eq:Ov}
\end{equation}
where $\Delta=\sqrt{1-r^2}$. The quadratic form $(\eta,\cO_v\eta)/2$ is restricted to $(\phi,\eta)=(\chi,\eta)=0$, with
\begin{equation}
 \phi=\dot q_r,
 \qquad N_T=(\phi,\phi),
 \qquad C'=(\chi,\psi_r).
\end{equation}
In particular, $N_T=(\phi,\phi)=\int\dot q_r^{\,2}d\tau$ is the holomorphic
norm of the translational zero mode.  It normalizes the translation
projector and supplies the collective-coordinate Jacobian
$J_T=\sqrt{N_T/(2\pi h)}$.

Now, differentiating the constrained-family identities with respect to $\tau_0$ and $r$ gives
\begin{equation}
 \cO_v\phi=0,
 \qquad
 \cO_v\psi_r=\kappa'(r)\chi
 =\frac{6r}{\Omega^2}\chi.
 \label{eq:tangentidentities}
\end{equation}
Let $G_T$ denote the inverse with translations removed,
\begin{equation}
 \begin{aligned}
 \mathcal O_{v,\tau}G_T(\tau,\tau')
 &=
 \delta(\tau-\tau')
 -\frac{\phi(\tau)\phi(\tau')}{N_T},
 \\[2pt]
 \int d\tau\,\phi(\tau)G_T(\tau,\tau')
 &=0.
 \end{aligned}
 \label{eq:GTdef}
\end{equation}
Removing translation alone does not enforce fixed separation.
The fluctuations must also satisfy $(\chi,\eta)=0$, which keeps
them tangent to the constraint surface.
Define the normal susceptibility $K=(\chi,G_T\chi)=\Omega^2C'/(6r)$.  Equation~\eqref{eq:tangentidentities} then gives $G_T\chi=(\Omega^2/6r)\psi_r$.  Projecting the translationally reduced propagator onto $(\chi,\eta)=0$ gives the propagator used in every fixed-$r$ diagram:
\begin{equation}
 \begin{aligned}
 G_\perp(\tau,\tau')
 &=G_T(\tau,\tau')
 -\frac{(G_T\chi)(\tau)(\chi G_T)(\tau')}{K}\\
 &=G_T(\tau,\tau')
 -\frac{\Omega^2}{6rC'}\psi_r(\tau)\psi_r(\tau').
 \end{aligned}
 \label{eq:Gperp}
\end{equation}
It obeys
\begin{equation}
 \int\dd\tau\,\phi(\tau)G_\perp(\tau,\tau')=0,
 \qquad
 \int\dd\tau\,\chi(\tau)G_\perp(\tau,\tau')=0.
 \label{eq:Gidentities}
\end{equation}
As a result, this $G_\perp$ is transverse to both collective directions.

The translation-reduced determinant is
\cite{GelfandYaglom,McKaneTarlie,DunneDeterminants}
\begin{equation}
 \frac{\det{}'\cO_v}{\det\cO_0}
 =-\frac{rN_T}{4\Omega^5},
 \qquad \cO_0=-\partial_\tau^2+\Omega^2.
 \label{eq:det}
\end{equation}
The prime excludes the translational zero mode, with the center fixed
by $(q-q_r,\phi)=0$. For real $r>0$, the identity
\[
 1=\int_0^\infty \dd r\,(-C'(r))\,
 \delta\!\left[\mathcal C[q]-\mathcal C[q_r]\right]
\]
introduces the separation coordinate and its Jacobian. The upper and
lower lateral measures follow by analytically continuing $q_r$ and
$C'(r)$ along the corresponding contours in the complex $r$ plane
\cite{BehtashCritical,BehtashComplex,BehtashKamata2026}. Combining both Jacobians with the determinant and $K$ we obtain
\begin{equation}
 \dd\mu_v(r)=\frac{\sqrt{-6\Omega^3C'(r)}}{\pi h}\,\dd r.
 \label{eq:measure}
\end{equation}
We choose the positive square root in~\eqref{eq:measure} for real
$r>0$ and continue it along the two lateral contours.  The resulting
measures are complex conjugates
\cite{BehtashComplex,BehtashCritical}.  With $r^2=-4hy$, the leading
measure becomes
\begin{equation}
 \dd\mu_v\,e^{-s/h}\longrightarrow
 -\frac{e^{-1/(2h)}}{\sqrt2\,\pi}e^{-y}\,\dd y
 \label{eq:leadingmeasure}
\end{equation}
for each mirror bion. The large-Euclidean-time vacuum amplitude fixes the sign, and with the
chosen lateral orientation, each mirror bion gives a positive local
energy contribution \cite{Coleman}.  Parity makes the contributions
from $q_r$ and $-q_r$ equal.  Their sum defines the local median
amplitude in the neutral-bion sector
$[\mathcal I\bar{\mathcal I}]$,
\begin{equation}
 \mathcal A^{\rm med}_{[\mathcal I\bar{\mathcal I}],\,\rm 1L}
 =
 \frac{\sqrt{2}}{\pi}\,
 e^{-1/(2h)}
 \left[1+O(h)\right].
 \label{eq:oneL}
\end{equation}
The relative $O(h)$ term is the two-loop correction evaluated below;
the global thimble multiplier is determined separately.

\subsection{Two-loop graphs}

\begin{figure*}[t]
 \centering
 \includegraphics[width=0.97\textwidth]{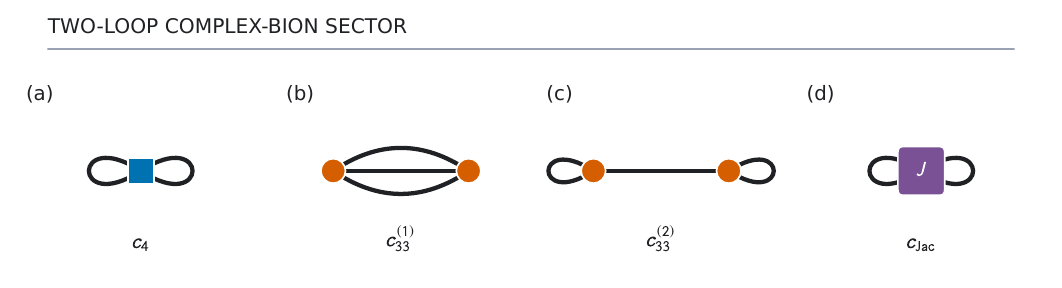}
 \caption{Connected two-loop structures on the constrained complex background. Every line denotes $G_\perp$; orange circles and blue squares denote cubic and quartic background vertices. Panel (d) represents the combined translation Faddeev--Popov, constraint/coarea, and curvature vertices entering ~\eqref{eq:cJac} \cite{Coleman,AffleckConstrained,AoyamaValleysQM}.}
\label{fig:biongraphs}
\end{figure*}

At relative order $h$, the fixed-$r$ contribution consists of the
quartic, theta, dumbbell, and collective-coordinate graphs shown in Figure~\ref{fig:biongraphs}. Vacuum subtraction and the separation integral are added
below. The local vertices follow from $s-\kappa(r)\cC$:
\begin{equation}
 \widetilde g_3=60\frac{\Delta^2}{\Omega^2}q_r^3-24q_r,
 \qquad
 \widetilde g_4=180\frac{\Delta^2}{\Omega^2}q_r^2-24,
 \label{eq:g34}
\end{equation}
with $V_3=\int\widetilde g_3\eta^3\dd\tau/6$ and
$V_4=\int\widetilde g_4\eta^4\dd\tau/24$.  Expanding
$\exp[-\sqrt h V_3-hV_4]$ gives
\begin{align}
 c_4(r)={}&-\frac18\int\dd\tau\,
 \widetilde g_4(\tau)G_\perp(\tau,\tau)^2,
 \label{eq:c4}\\
 c_{33}^{(1)}(r)={}&\frac1{12}\iint\dd\tau\dd\tau'\,
 \widetilde g_3(\tau)\widetilde g_3(\tau')
 G_\perp(\tau,\tau')^3,
 \label{eq:c331}\\
 c_{33}^{(2)}(r)={}&\frac18\iint\dd\tau\dd\tau'\,
 \widetilde g_3(\tau)\widetilde g_3(\tau')
 G_\perp(\tau,\tau')\nonumber\\
 &\hspace{9mm}\times G_\perp(\tau,\tau)
 G_\perp(\tau',\tau').
 \label{eq:c332}
\end{align}

Fluctuations also change the collective-coordinate measure and
therefore contribute at the same loop order as the interaction
graphs. Solving the nonlinear fixed-$C$ condition and expanding
the translation and constraint Jacobians gives \cite{Coleman,AffleckConstrained,AoyamaValleysQM}
\begin{equation}
 c_{\rm Jac}(r)=\left\langle j_2-j_1V_3
 +\frac14\mathcal Q\mathcal R-\frac{\mathcal Q^2}{8K}\right\rangle_{G_\perp}.
 \label{eq:cJac}
\end{equation}
Equation~\eqref{eq:cJac} combines the translation and fixed-separation
Jacobians with the curvature of the constraint surface.
Appendix~\ref{app:collective} gives the corresponding Green-function
expressions. The last two terms arise from the curvature of the constraint surface.

To isolate the fluctuation correction associated with the bion,
we subtract the vacuum graphs using the same infrared regulator, as in real-instanton loop calculations \cite{EscobarQuartic,EscobarSine}.  At $q=0$ the cubic vertex vanishes, so the only order-$h$ vacuum graph is the quartic figure-eight.  For the Dirichlet inverse in a symmetric box $[-L,L]$,
\begin{equation}
 \begin{aligned}
 c_{\rm vac}^{D}(L)
 &=-\frac18\int_{-L}^{L}\dd\tau
 (-24)G_0^D(\tau,\tau)^2\\
 &=\frac{3L}{2\Omega^2}-\frac{9}{8\Omega^3}
 +O\!\left(Le^{-4\Omega L}\right).
 \end{aligned}
 \label{eq:cvac}
\end{equation}
The extensive pieces enter the bion coefficient through the finite combination $c_4-c_{\rm vac}^{D}(L)$ in~\eqref{eq:c1lat}.

\subsection{Lateral separation integral and final sum}

For $q_r^{(\sigma)}=\sigma q_r$, let $\Gamma_\pm$ denote the two
lateral $r$ cycles. Near the dilute endpoint,
\[
r_\pm^2=-4hy\pm i0,\qquad y>0,
\]
and the saddle $r^2=-3h$ lies at $y=3/4$. The half-sum of the
conjugate cycles defines the local median amplitude, while the sum
over $\sigma=\pm1$ includes excursions through both outer wells
\cite{BogomolnyIbarI,ZinnJustinMulti,DunneTriple,
BehtashComplex,BehtashKamata2026}.

With the isolated $e^{-S_I/h}$ term absent, the two mirror bions generate the one-bion sector at order $e^{-1/(2h)}$; the next sector begins at $e^{-1/h}$.  Their leading normalized separation weight is $e^{-y}$ \cite{DunneTriple,KamataExactWKB}.  For a fixed-$r$ quantity $F$ define
\begin{equation}
 \langle F\rangle_{0,\pm}=\int_0^\infty e^{-y}
 F[r_\pm(y)]\,\dd y.
 \label{eq:cycleavg}
\end{equation}
The $O(h)$ expansion of the exact action and the measure~\eqref{eq:measure} produces the explicit quasi-zero-mode (QZM) term
\begin{equation}
 c_{\rm QZM}^{\pm}(h)=\frac{103}{16}
 +\frac38\left(\ln h-\gamma\pm i\pi\right)\,,
 \label{eq:cqzm}
\end{equation}
with $\gamma$ the Euler–Mascheroni constant.
This term follows by expanding the exact action and measure along each lateral continuation.  Its logarithm and ambiguity cancel in ~\eqref{eq:c1lat} \cite{DunneUnsalZero,BehtashCritical,AoyamaValleysQM}.

The complete lateral coefficient is
\begin{equation}
 \begin{aligned}
 c_1^\pm(h;L)={}&c_{\rm QZM}^\pm(h)
 +\left\langle c_4+c_{33}^{(1)}+c_{33}^{(2)}\right.\\
 &\left.\hspace{10mm}+c_{\rm Jac}-c_{\rm vac}^{D}(L)\right\rangle_{0,\pm}.
 \end{aligned}
 \label{eq:c1lat}
\end{equation}
The two lateral coefficients obey $c_1^-=(c_1^+)^*$.  We therefore define the local median coefficient by the standard lateral prescription \cite{BehtashComplex,BehtashCritical,KamataExactWKB}:
\begin{equation}
 \med c_1\equiv\frac{c_1^++c_1^-}{2}
 =\operatorname{Re}c_1^+ .
 \label{eq:median}
\end{equation}
Table~\ref{tab:individual} shows how the separate contributions combine at finite $h$. After continuum extrapolation, its entries give
$c_1^+=-0.78015+0.03343i$. Taking the median removes the imaginary
part, but the remaining real value still contains finite-$h$
and finite-box effects. The infinite-box and $h\to0$
extrapolations described below isolate the asymptotic two-loop
coefficient.

\begin{table}[!t]
\centering
\caption{Upper-lateral decomposition of ~\eqref{eq:c1lat} at $h=2\times10^{-5}$ and $L=12$. Each term is extrapolated to the continuum using the same 80-node quadrature. The entries are rounded independently and reproduce the total within rounding.}
\label{tab:individual}
\begingroup
\renewcommand{\arraystretch}{1.30}
\setlength{\tabcolsep}{7pt}
\begin{tabular}{@{}lrr@{}}
\toprule
Term & \multicolumn{1}{c}{Real part} &
\multicolumn{1}{c}{Imaginary part} \\
\midrule
$\langle c_4\rangle_{0,+}$          & $-16.73313$ & $ 20.03052$ \\
$\langle c_{33}^{(1)}\rangle_{0,+}$ & $ 15.01678$ & $-11.39025$ \\
$\langle c_{33}^{(2)}\rangle_{0,+}$ & $ 29.21665$ & $-14.20995$ \\
\addlinespace[2pt]
$\langle c_{\rm Jac}\rangle_{0,+}$  & $-13.57006$ & $  4.42501$ \\
$-c_{\rm vac}^{D}(L)$                & $-16.87402$ & $  0$       \\
\addlinespace[2pt]
$c_{\rm QZM}^{+}(h)$                 & $  2.16363$ & $  1.17810$ \\
\midrule
$c_1^+(h;L)$                         & $-0.78015$  & $ 0.03343$  \\
\bottomrule
\end{tabular}
\endgroup
\end{table}

The dilute limit is taken before the lateral integral. Let
$B(r;h,L)$ denote the vacuum-subtracted fixed-$r$ graph sum in
~\eqref{eq:c1lat}, and define
\begin{equation}
 B_0(r)=\lim_{h\to0}\lim_{L\to\infty}B(r;h,L).
\end{equation}
For the complete sum, the shift $\tau\to\tau+i\pi/(2\Omega)$
relates $q_r$ to $q_{-r}$. Once the background and Jacobian
terms are combined, the odd powers of $r$ cancel. Hence
\begin{equation}
 \begin{aligned}
 B_0(r)={}&-\frac34\ln r+C_0
 +r^2\left(a_2\ln^2r+a_1\ln r+a_0\right)+\cdots,\\
 c_{\rm 2L}(r)={}&B_0(r)+\frac34\ln r
 -\frac34\ln2+\frac{103}{16},\\
 c_{\rm 2L}={}&\lim_{r\to0^+}c_{\rm 2L}(r).
 \end{aligned}
 \label{eq:dilute}
\end{equation}
On the upper contour, $r_+=2i\sqrt{hy}$, the logarithm
generates the endpoint term $-(3/8)\ln y$, which is integrated
analytically. No exact-WKB input enters this extraction.

\begin{figure}[!t]
 \centering
 \includegraphics[width=0.92\columnwidth]{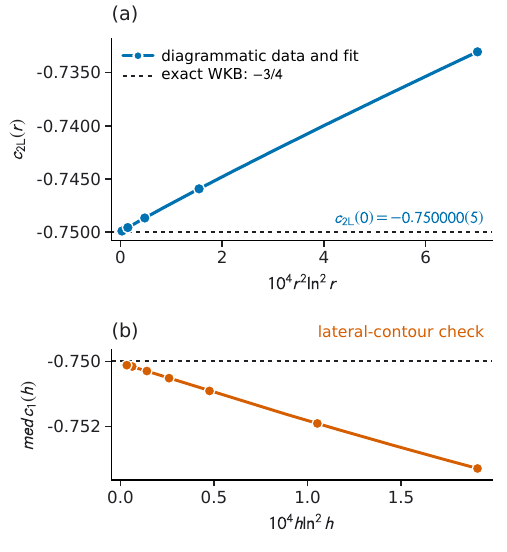}
 \caption{Two-loop convergence after vacuum subtraction and continuum extrapolation. (a)~Real-family estimator $c_{\rm 2L}(r)$ defined in ~\eqref{eq:dilute}; the solid curve is its dilute-limit fit. (b)~Endpoint-subtracted median lateral sum fitted to $c_0+h(\alpha\ln^2h+\beta\ln h+\gamma_0)$. The points are the numerical data, and the dashed lines mark the exact-WKB value $-3/4$, which is not used in either fit.}
 \label{fig:twoLoopConvergence}
\end{figure}

Continuum and infinite-box extrapolations, independently checked with
a sine--Galerkin discretization, yield $-0.74999982$ and
$-0.74999998$ from the pointwise and dilute-limit analyses,
respectively. Appendix~\ref{app:numerics} describes the numerical
procedure. At fixed $h$, we first take the continuum limit at fixed
$L$ and then the infinite-box limit:
\begin{equation}
 c_1^\pm(h)\equiv
 \lim_{L\to\infty}\lim_{\Delta\tau\to0}
 c_1^\pm(h;L,\Delta\tau),
 \label{eq:continuumlimit}
\end{equation}
where $c_1^\pm(h;L,\Delta\tau)$ is the finite-grid evaluation of
~\eqref{eq:c1lat} along the two lateral contours.

The median dilute limit gives
\begin{equation}
 c_{\rm 2L}^{\rm diag}
 \equiv\lim_{h\to0}\med c_1(h)
 =-0.750000(5).
 \label{eq:finalsum}
\end{equation}
Here $c_1$ denotes the complete graph sum in ~\eqref{eq:c1lat}.
The quoted uncertainty accounts for the spread under the continuum,
box-size, and regulator extrapolations. Combining
Eqs.~\eqref{eq:oneL} and \eqref{eq:finalsum}, we obtain the local
median one-bion amplitude through relative order $h$,
\begin{equation}
 \mathcal A^{\rm med}_{[I\bar I]}
 =
 \frac{\sqrt{2}}{\pi}\,e^{-1/(2h)}
 \left[1-0.750000(5)h+o(h)\right].
 \label{eq:finalamplitude}
\end{equation}
Both the leading prefactor and its first fluctuation correction have been obtained diagrammatically. The corresponding spectral
contribution is
\begin{equation}
 \delta E_{[I\bar I]}
 =
 \mathfrak n_{[I\bar I]}\,
 \mathcal A^{\rm med}_{[I\bar I]},
\end{equation}
where \(\mathfrak n_{[I\bar I]}\) is the global thimble multiplier.
\subsection{Exact-WKB check}

The median exact-WKB quantization condition supplies an independent
determination of the physical one-bion contribution. Specializing it to
$p=1$, $k=0$, and $\hbar_K=h/2$ gives
\cite{KamataExactWKB}
\begin{equation}
 \delta E^{\rm WKB}_{[I\bar I]}
 =
 \frac{\sqrt{2}}{\pi}\,e^{-1/(2h)}
 \left[
 1-\frac34h+O(h^2)
 \right].
 \label{eq:wkbcheck}
\end{equation}
Appendix~\ref{app:wkb} gives the normalization map. Comparison with
~\eqref{eq:finalamplitude} confirms the diagrammatic coefficient
$c_{\rm 2L}^{\rm diag}=-3/4$ and fixes
$\mathfrak n_{[I\bar I]}=1$ in the present normalization. Neither the
exact-WKB coefficient nor this multiplier enters the diagrammatic
extraction.

Multi-bion sectors begin at
$O(e^{-1/h}[1+|\ln h|])$
\cite{DunneTriple,KamataExactWKB}. Equations~\eqref{eq:finalamplitude}
and \eqref{eq:wkbcheck} therefore isolate the one-bion sector.

\section{Conclusions}
\label{conclusions}

Factorization fixes $E_0=\mu/2$ at positive coupling and forces all
perturbative coefficients to vanish. Figures~\ref{fig:P12} and
\ref{fig:P3} verify this cancellation directly through $O(\nu^3)$.
After continuation to $\nu=-g$, the formal zero mode becomes
nonnormalizable. The factorized Hamiltonian remains positive, but its
ground state is lifted nonperturbatively above $\mu/2$.

The unequal central and outer curvatures suppress the normalized
one-instanton contribution. The leading local semiclassical amplitude
is then supplied by a complex neutral bion. We computed its projected
propagator, one-loop measure, and complete vacuum-subtracted two-loop
graph sum, including the non-Gaussian separation integral. The result is
\[
 c_{\rm 2L}^{\rm diag}=-0.750000(5)\,.
\]
The diagrammatic coefficient agrees with the exact-WKB value $-3/4$.
Exact WKB also fixes the global multiplier
$\mathfrak n_{[I\bar I]}=1$ in the present normalization. This establishes a direct diagrammatic calculation of the local
median one-bion amplitude.

\begin{acknowledgments}
The author warmly thanks E. Shuryak and A. V. Turbiner for introducing him to instanton calculus in quantum mechanics and for the many illuminating discussions during their earlier collaborations.
\end{acknowledgments}

\section*{Data Availability Statement}

No external data were used. All numerical results follow from the
equations and procedures described in this article.

\appendix

\section{Wick factors in the perturbative sector}
\label{app:wick}

For a connected Gaussian graph with vertex valences $d_i$, let $l_i$ be the number of self-contractions and $m_{ij}$ the number of lines between vertices.  The valence constraints and multiplicity are
\begin{equation}
 2l_i+\sum_{j\ne i}m_{ij}=d_i,
 \qquad
 N_G=\frac{\prod_i d_i!}
 {\prod_i2^{l_i}l_i!\prod_{i<j}m_{ij}!}.
 \label{eq:wick}
\end{equation}
For two vertices,
$\int_{-\infty}^{\infty}e^{-m|\tau|}\dd\tau=2/m$.
For three vertices, fixing one time at zero and splitting the remaining plane into its six orderings gives
\begin{equation}
 I(a,b,c)=\sum_{\pi\in S_3}
 \frac{1}{\kappa_1(\pi)\kappa_2(\pi)},
\end{equation}
where $\kappa_k$ is the number of lines crossing the $k$th cut of the ordering.  These two formulas reproduce all rational entries in Figs.~\ref{fig:P12} and \ref{fig:P3}.

\section{Constrained-family determinant, measure, and separation}
\label{app:qzm}

For $r\ne0$ the translation-projected inverse has the finite-part resolvent representation \cite{McKaneTarlie,DunneDeterminants}
\begin{equation}
 G_T(\tau,\tau')=\underset{\lambda=0}{\operatorname{FP}}
 \left[-\frac{f_{\lambda,-}(\tau_<)
 f_{\lambda,+}(\tau_>)}{W_\lambda}\right].
 \label{eq:GT}
\end{equation}
Here $f_{\lambda,-}$ and $f_{\lambda,+}$ decay at $-\infty$ and $+\infty$, respectively,
$W_\lambda=f_{\lambda,-}f'_{\lambda,+}-f'_{\lambda,-}f_{\lambda,+}$, and FP removes the simple translation pole.  The combined expression in ~\eqref{eq:Gperp} is taken before the $r\to0$ limit.

Define
\begin{equation}
 \ell=\operatorname{arcosh}(1/r)
 =\ln\frac{1+\Delta}{r}.
\end{equation}
We take $\Delta\to1$ as $r\to0$ and begin with the principal logarithm for $\operatorname{Re}r>0$.  Continuous transport above or below the negative $r^2$ axis defines the two lateral branches \cite{BehtashComplex,BehtashCritical}.  Direct integration gives
\begin{align}
 N_T&=\frac{\Omega^3(\Delta-r^2\ell)}{2\Delta^3},\\
 C'&=\frac{\Omega^5}{12\Delta^7}
 \left[3r(4+r^2)\ell-\frac{2\Delta^3}{r}-15r\Delta\right].
 \label{eq:NTCprime}
\end{align}
The translation mode has tails $\phi\sim A_-e^{\Omega u}$ and
$\phi\sim A_+e^{-\Omega u}$, with
$A_-=-A_+=\Omega^2\sqrt{2/r}$.  Therefore
\begin{equation}
 \frac{\det{}'\cO_v}{\det\cO_0}
 =\frac{N_T}{2\Omega A_-A_+}
 =-\frac{rN_T}{4\Omega^5},
\end{equation}
which proves ~\eqref{eq:det} directly from the zero-mode tails \cite{McKaneTarlie,DunneDeterminants}.

The one-loop measure combines the translation Jacobian, the fixed-$\cC$ coarea factor, and the primed determinant \cite{Coleman,AffleckConstrained,AoyamaValleysQM}:
\begin{equation}
 \sqrt{\frac{N_T}{2\pi h}}
 \frac{C'}{\sqrt{2\pi hK}}
 \left(\frac{\det{}'\cO_v}{\det\cO_0}\right)^{-1/2}
 =\frac{\sqrt{-6\Omega^3C'}}{\pi h}.
 \label{eq:measureproduct}
\end{equation}
This establishes ~\eqref{eq:measure}; its phase uses the branch obtained by analytic continuation from real $r>0$ along the chosen lateral path \cite{BehtashComplex,BehtashCritical}.

For the separation expansion, let $I_n=\int q_r^n\dd\tau$.  The moments needed here are
\begin{align}
 I_2&=\frac{\Omega\ell}{\Delta},
 &I_4&=\frac{\Omega^3(\ell-\Delta)}{\Delta^3},\\
 I_6&=\frac{\Omega^5}{2\Delta^5}
 \left[(2+r^2)\ell-3\Delta\right],
 &C(r)&=\frac{I_6}{6}.
\end{align}
They give the exact constrained-family action
\begin{equation}
 s(r)=\Omega^2I_2-2I_4
 +\frac{\Omega^2+\Delta^2}{2\Omega^2}I_6.
 \label{eq:exactaction}
\end{equation}
This follows from
$\dot q_r^{\,2}=\Omega^2q_r^2-2q_r^4+\Delta^2q_r^6/\Omega^2$
and obeys $s'=\kappa C'$ as a check.  Set
$r^2=hz$ and $L_z=\ln[2/\sqrt{hz}]$.  At fixed $z$,
\begin{align}
 \frac{s(r)}{h}
 &=\frac{1}{2h}+\frac32L_z-\frac z4
 +h\mathcal B_1(z,L_z)+O(h^2L_z),\\
 \mathcal B_1(z,L_z)
 &=\left(\frac34z^2+\frac92z+\frac{27}{4}\right)L_z
 -z^2-6z-\frac{135}{16},\\
 rC'(r)&=-\frac16
 +h\left(zL_z-\frac{19}{12}z-\frac54\right)
 +O(h^2L_z).
 \label{eq:smallrexpansions}
\end{align}
Combining these expansions with the determinant and Jacobians in ~\eqref{eq:measureproduct} gives the normalized lateral weight
\begin{equation}
 e^{-y}\left[1+hF_1(-4y,L_z)+O(h^2L_z^2)\right],
\end{equation}
where
\begin{equation}
 F_1=z^2+\frac{43}{4}z+\frac{231}{16}
 -L_z\left(\frac34z^2+\frac{15}{2}z+\frac{27}{4}\right).
\end{equation}
On $z=-4y\pm i0$,
$L_z=-\tfrac12\ln(hy)\mp i\pi/2$.  Using
$\int_0^\infty y^ne^{-y}\dd y=n!$ and
$\int_0^\infty e^{-y}\ln y\,\dd y=-\gamma$
then gives ~\eqref{eq:cqzm}, including its logarithm and lateral sign.

\section{Collective-coordinate graphs}
\label{app:collective}

The fixed-constraint and collective-coordinate expansion follows the standard framework of Refs.~\cite{Coleman,AffleckConstrained,AoyamaValleysQM,EscobarQuartic}.  Let $\bar\psi=\psi_r/C'$ and define
\begin{align}
 C_2[f,g]&=\int5q_r^4fg\,\dd\tau,\\
 C_3[f,g,k]&=\int20q_r^3fgk\,\dd\tau,\\
 \mathcal Q&=C_2[\eta,\eta],\\
 \mathcal R&=\int\widetilde g_3\eta^2\bar\psi\,\dd\tau,\\
 K&=(\chi,G_T\chi)=\frac{\Omega^2C'}{6r}.
\end{align}
The fixed-constraint surface is
\begin{equation}
 q=q_r+\sqrt h\,\eta-\frac h2\mathcal Q\bar\psi+O(h^{3/2}),
 \qquad (\phi,\eta)=(\chi,\eta)=0.
\end{equation}
Its curvature contributes $-\mathcal Q\mathcal R/4+\mathcal Q^2/(8K)$ to the quartic interaction.

The constraint and translation Jacobians are
\begin{align}
 \frac{J_C}{J_C[0]}={}&1-\sqrt h B_1
 +h\left(B_1^2+\frac12\mathcal QD_2-\frac12E\right)+O(h^{3/2}),\\
 \frac{J_T}{J_T[0]}={}&1-\sqrt h D_1
 +\frac h2\mathcal QD_0+O(h^{3/2}),
\end{align}
where
\begin{align}
 B_1&=C_2[\eta,\bar\psi],
 &D_2&=C_2[\bar\psi,\bar\psi],\\
 E&=C_3[\eta,\eta,\bar\psi],
 &D_1&=\frac{(\eta,\ddot q_r)}{N_T},\\
 D_0&=\frac{(\bar\psi,\ddot q_r)}{N_T}.
\end{align}
Thus $j_1=-(B_1+D_1)$ and
\begin{equation}
 j_2=B_1^2+\frac12\mathcal Q(D_2+D_0)-\frac12E+B_1D_1,
\end{equation}
which gives ~\eqref{eq:cJac} by Wick contraction with $G_\perp$.

For the explicit contractions, define
\begin{equation}
 \begin{aligned}
 a&=5q_r^4,
 &\beta&=a\bar\psi,\\
 d&=\ddot q_r/N_T,
 &e&=20q_r^3\bar\psi,\\
 b&=\widetilde g_3\bar\psi.
 \end{aligned}
\end{equation}
Writing $G=G_\perp$, the explicit contracted expression is
\begin{align}
 c_{\rm Jac}={}&\iint\beta_xG_{xy}\beta_y
 +\frac{D_2+D_0}{2}\int a_xG_{xx}
 \nonumber\\
 &-\frac12\int e_xG_{xx}
 +\iint\beta_xG_{xy}d_y
 \nonumber\\
 &+\frac12\iint(\beta_x+d_x)\widetilde g_{3,y}
 G_{xy}G_{yy}
 \nonumber\\
 &+\frac14\iint a_xb_y
 \left(G_{xx}G_{yy}+2G_{xy}^2\right)
 \nonumber\\
 &-\frac1{8K}\iint a_xa_y
 \left(G_{xx}G_{yy}+2G_{xy}^2\right).
 \label{eq:cJacExplicit}
\end{align}
All integrals in ~\eqref{eq:cJacExplicit} use the same box as ~\eqref{eq:cvac}.

\section{Exact-WKB normalization}
\label{app:wkb}

In the notation of Ref.~\cite{KamataExactWKB},
\begin{equation}
 \begin{aligned}
 \widehat H_K={}&-\frac{\hbar_K^2}{2}\partial_q^2
 +\frac12\mathcal W'^2+\frac{p\hbar_K}{2}\mathcal W'',\\
 \mathcal W'={}&\frac12q(q^2-1),
 \qquad \hbar_K=\frac h2,\quad p=1,\\
 \cH_+(h)={}&4\widehat H_K,
 \qquad \delta E=2\varepsilon_K,
 \quad \varepsilon_K=E_K/\hbar_K.
 \end{aligned}
 \label{eq:wkbmap}
\end{equation}
The deformation label is $\mathfrak q_K=(3p-1)/2=1$, and the even central ground state has $k=0$.  Set $\xi=-\bar F$, with $\bar F$ the central-cycle spectral coordinate.  The even median factor and its bion cycle are
\begin{align}
 f_+&=2i\left[-1+\sqrt{1+\bar{\mathfrak B}}
 \cos(2\pi\xi)\right],\nonumber\\
 \bar{\mathfrak B}
 &=\frac{2\pi e^{-G_K(-\xi,\hbar_K)}
 2^{1/2-\xi}\hbar_K^{-3\xi}}
 {\Gamma(\xi)\Gamma(1+2\xi)}\nonumber\\
 &=2\sqrt2\pi\xi\mathcal B_0
 \left[1+O\!\left(\xi[1+|\ln\hbar_K|]\right)\right],
 \label{eq:wkbbioncycle}
\end{align}
where $\mathcal B_0=e^{-G_K(0,\hbar_K)}$.  The last line uses $1/\Gamma(\xi)=\xi+O(\xi^2)$.  Expanding $f_+=0$ yields
\begin{equation}
 \begin{aligned}
 -4\pi^2\xi^2+2\sqrt2\pi\mathcal B_0\xi+\cdots&=0,\\
 \xi_{\rm phys}&=\frac{\mathcal B_0}{\sqrt2\pi}
 +O\!\left(\mathcal B_0^2[1+|\ln\hbar_K|]\right).
 \end{aligned}
 \label{eq:wkbroot}
\end{equation}
The zero root is the formal nonnormalizable factorized solution.  For the physical root,
\begin{equation}
 \begin{aligned}
 \mathcal B_0&=e^{-1/(4\hbar_K)}
 \left[1-\frac32\hbar_K+O(\hbar_K^2)\right],\\
 \varepsilon_K&=\xi+O(\hbar_K^2\xi,\hbar_K\xi^2).
 \end{aligned}
\end{equation}
Together with ~\eqref{eq:wkbmap}, this gives ~\eqref{eq:wkbcheck}.

\vspace{-0.5cm}

\section{Numerical definition of the regulated sum}
\label{app:numerics}

For the Dirichlet vacuum kernel on $[-L,L]$, the subtraction entering the graph sum is
\begin{equation}
 \begin{aligned}
 c_{\rm vac}^{D}(L)
 ={}&\frac{3L}{2\Omega^2}
 +\frac{9L}{4\Omega^2\sinh^2(2\Omega L)}
 -\frac{9}{8\Omega^3}\coth(2\Omega L).
 \end{aligned}
 \label{eq:cvacexact}
\end{equation}
Its large-$L$ limit gives ~\eqref{eq:cvac}. The finite-$L$ expression must be retained because the background and vacuum graphs are subtracted before taking the infinite-box limit.

We discretize $[-L,L]$ with spacing $\Delta\tau$ and impose Dirichlet boundary conditions. The projected covariance is the upper-left block of the bordered inverse
\begin{equation}
 \begin{pmatrix}
 \Delta\tau\,\cO_v & \Delta\tau\,Z\\
 \Delta\tau\,Z^{\mathsf T} & 0
 \end{pmatrix}^{-1},
 \qquad
 Z=[\boldsymbol\phi\ \boldsymbol\chi].
 \label{eq:KKT}
\end{equation}
This construction imposes orthogonality to the translational mode $\boldsymbol\phi$ and to the constraint normal $\boldsymbol\chi$ directly, without subtracting nearly singular eigenvalues. The vacuum inverse is computed on the same grid. Equations~\eqref{eq:c4}--\eqref{eq:cvac} and~\eqref{eq:cJacExplicit} then reduce to matrix contractions. The resulting finite-grid approximation to ~\eqref{eq:c1lat} is denoted by
$c_1^\pm(h;L,\Delta\tau)$.

For Table~\ref{tab:individual}, we use $L=12$, the three grids
$N=321,481,641$, and an 80-node Gauss--Laguerre quadrature for integrals of the form
$
 \int_0^\infty e^{-y}F(y)\,\dd y.
$
The same nodes and weights are used for every term so that the displayed contributions sum consistently. In the total, the endpoint singularity $-(3/8)\ln y$ is subtracted before quadrature and its exact integral, $3\gamma/8$, is restored analytically. Relative to the unsubtracted $n$-node rule, this changes the result by
\begin{equation}
 \frac{3}{8}
 \left(\sum_{i=1}^{n}w_i\ln y_i+\gamma\right),
\end{equation}
which equals $0.00295582493$ for $n=80$. After endpoint subtraction, increasing the quadrature order from 24 to 64 nodes changes the real part of the total by
$2.7\times10^{-6}$ at $h=10^{-6}$. Equation~\eqref{eq:dilute} gives the finite-$h$ form
\[
 c_0+h\left(\alpha\ln^2h+\beta\ln h+\gamma_0\right)+\cdots;
\]
the complete sum contains no term proportional to $\sqrt h$.

Figure~\ref{fig:twoLoopConvergence} uses $L=18$ and
$N=721,1081,1441$, followed by a quadratic extrapolation in
$(\Delta\tau)^2$. Changing the fit basis and the $r$ window shifts the result by less than $3\times10^{-7}$. An independent Dirichlet sine--Galerkin calculation with up to 1440 modes, performed at $L=18$ and $20$, agrees within $4.4\times10^{-7}$. Variations of the cutoff model remain below $3\times10^{-6}$. These tests estimate the continuum and infinite-box limits in ~\eqref{eq:continuumlimit} and lead to the conservative uncertainty
$5\times10^{-6}$ quoted in ~\eqref{eq:finalsum}.

\bibliography{qes_sextic_refs}

\end{document}